\documentclass[12pt,preprint]{aastex}

\def\ltsima{$\; \buildrel < \over \sim \;$}
\def\gtsima{$\; \buildrel > \over \sim \;$}
\def\lsim{\lower.5ex\hbox{\ltsima}}
\def\gsim{\lower.5ex\hbox{\gtsima}}
\def\lapp{\ifmmode\stackrel{<}{_{\sim}}\else$\stackrel{<}{_{\sim}}$\fi}
\def\gapp{\ifmmode\stackrel{>}{_{\sim}}\else$\stackrel{<}{_{\sim}}$\fi}

\newdimen\minuswidth    %define @ width of minus sign for tables
\setbox0=\hbox{$-$}
\minuswidth=\wd0
\catcode`@=\active
\def@{\kern\minuswidth}
\setbox0=\hbox{\rm0}
\shorttitle{Blue Stragglers in M2}
\shortauthors{Dalessandro et al.}
 
\begin{document} 
\title{Multiwavelength photometry in the Globular Cluster M2\footnote{Based on observations with the
NASA/ESA {\it HST} (Prop. 8709 and Prop.10775), obtained at the
Space Telescope Science Institute, which is operated by AURA, Inc., under
NASA contract NAS5-26555. Also based on EMMI observations (Prop 079.D-0325) collected
at the European Southern Observatory, La Silla, Chile. Based on observations with MegaPrime/MegaCam, a joint project 
of CFHT and CEA/DAPNIA, at the Canada-France-Hawaii Telescope (CFHT), which is operated by the National Research Council 
(NRC) of Canada, the Institute National des Sciences de l'Univers of the Centre National de la Recherche Scientifique of France, 
and the University of Hawaii. It is also based on GALEX observations
(program GI-056)}.}

\author{
E. Dalessandro\altaffilmark{2,3,4},
G. Beccari\altaffilmark{5},
B. Lanzoni\altaffilmark{2},
F.R. Ferraro\altaffilmark{2}, 
R. Schiavon\altaffilmark{6},
R.T. Rood\altaffilmark{7}}

\affil{\altaffilmark{2} Dipartimento di Astronomia, Universit\`a degli Studi
di Bologna, via Ranzani 1, I--40127 Bologna, Italy}
\affil{\altaffilmark{3} ASI,Centro di Geodesia Spaziale, contrada Terlecchia,
I-75100, Matera, Italy}   
\affil{\altaffilmark{4} INAF, Osservatorio Astronomico di Bologna, via Ranzani 1, 
I--40127 Bologna, Italy}   
\affil{\altaffilmark{5} European Space Agency, Space Science Department, Keplerlaan 1, 2200 AG Noordwijk, Netherlands} 
%\affil{\altaffilmark{3} INAF--Osservatorio Astronomico di Bologna, via
%Ranzani 1, I--40127 Bologna, Italy}  %, {\tt barbara.lanzoni@bo.astro.it}}
%
\affil{\altaffilmark{6} Gemini Observatory, 670 North A'ohoku Place, Hilo, HI 96720, USA}
\affil{\altaffilmark{7} Astronomy Department, University of Virginia,
P.O. Box 400325, Charlottesville, VA, 22904}

\date{Accepted by ApJS on March 19, 2009}

\begin{abstract}
We present a multiwavelength photometric analysis of the globular
cluster M2.  The data-set has been obtained by combining
high-resolution (HST/WFPC2 and ACS) and wide-field (GALEX) space
observations and ground based (MEGACAM-CFHT, EMMI-NTT) images.  The
photometric sample covers the entire cluster extension from the very
central regions up to the tidal radius and beyond. It allows an
accurate determination of the cluster center of gravity and other
structural parameters derived from the star count density
profile. Moreover we study the BSS population and its radial
distribution. A total of 123 BSS has been selected, and their radial
distribution has been found to be bimodal (highly peaked in the
center, decreasing at intermediate radii and rising outward), as
already found in a number of other clusters.  The radial position of
the minimum of the BSS distribution is consistent with the radius of
avoidance caused by the dynamical friction of massive ($1.2M_{\odot}$)
objects over the cluster age. We also searched for gradients in the
red giant branch (RGB) and the asymptotic giant branch (AGB)
populations. At the $2\sigma$ level we found an overabundance of AGB
stars within the core radius and confirmed the result of Sohn et
al.(1996) that the central region of M2 is bluer than the outer
part. We show that the latter is due to a deficit of very luminous RGB
stars in the central region.

\end{abstract} 


\section{INTRODUCTION}

In this paper we present multiwavelength observations of the Galactic
globular cluster M2 (NGC~7089).  These observations are part of a
large project aimed at characterizing the ultraviolet (UV) bright
populations of old stellar systems and determining the impact of
stellar dynamics on the cluster evolution by studying their
``exotic'' populations. As in our previous study of NGC~1904 (Lanzoni
et al.  2007b), we use HST high-resolution UV and optical data for the
high density central region of the cluster and a combination of
ground-based wide-field optical data (MEGACAM-CFHT and EMMI-NTT) and
UV data from the Galaxy Evolution Explorer (GALEX) for the cluster outskirts. 
Combining these
samples allows an accurate determination of the center of
gravity, the stellar density profile and the structural parameters. In this paper we focus on the
Blue Straggler Star (BSS) population as tracers of the dynamical state
of the host cluster and products of the interplay between stellar
evolution and stellar dynamics. We also discuss possible radial
gradients in the Asymptotic Giant Branch (AGB) stars
and other stellar populations. We defer a discussion of the Horizontal
Branch (HB) population to a future paper.

In the optical color-magnitude diagram (CMD) BSSs are bluer (hotter)
and brighter than the main-sequence (MS) stars, thus mimicking a
stellar population significantly younger than the ``normal'' cluster
stars.  As shown by Shara et al. (1997), BSSs are more massive than
normal stars, suggesting that some mass-increasing mechanism drives
their formation.  Possible explanations involve mass transfer between
binary companions, the merger of a binary system, and the collision
between single and/or binary stars (McCrea 1964; Zinn \& Searle
1976). Clear differentiation among these possibilities is difficult,
since primordial binaries can sink to the cluster center, where
stellar collisions may significantly alter their evolution. Similarly,
gravitational interactions can generate new binary systems and
possibly kick them out of the cluster core. With this caveat, we
define primordial binary BSS (PB-BSS) those formed by mass
transfer processes (possibly up to complete coalescence) in primordial
binaries which evolved in isolation in the cluster. Collisional BSS
(COL-BSS) are those generated by mechanisms where stellar collisions
played a major role. We therefore expect PB-BSS to mainly populate the
external regions of the cluster, where the collision probabilities are
lower.  COL-BSSs preferentially form in the central regions because of
the higher stellar densities \footnote{A distinction between PB-BSS
and COL-BSS requires high resolution spectroscopic studies (see the
case of 47~Tucane in Ferraro et al. 2006): in fact characteristic
chemical signatures are expected on the surface of PB-BSSs so that
accurate measurement of the stellar surface abundances can 
distinguish between the two types of stars (Sarna \& de Greve 1996),
while they are not predicted in the case of COL-BSSs (Lombardi et
al. 1995).}. These formation mechanisms may work simultaneously with
different efficiency depending on the environment (Fusi Pecci et al. 1993;
Ferraro et al. 1999, 2003; Bellazzini et al 2002).

Observed BSS radial distributions have been particularly important
in demonstrating the complex interplay of the various
phenomena. Typically the BSS radial distributions have been found to be
bimodal (peaked in the clusters center and outskirts and with a
dip at intermediate radii; see references in Dalessandro et
al. 2008a; see also Beccari et al. 2008 for M53).  Only two clusters
deviate from this pattern: $\omega$~Centauri ($\omega$~Cen, Ferraro et
al. 2006b) and NGC~2419 (Dalessandro et al. 2008b). In those clusters
the BSS radial distribution is indistinguishable from that of the
other cluster stars.  Simple dynamical simulations (Mapelli et
al. 2004, 2006; Lanzoni et al. 2007a) suggest that the observed
bimodality can be modelled assuming that PB-BSSs and COL-BSSs co-exist
in the same cluster with relative fractions that vary from one case to
another. The radial distributions observed in NGC~2419
and $\omega$\,Cen could be the observational evidence that
mass-segregation processes have played a minor role in altering the
BSS radial distributions and that the observed BSS population is
mainly composed of PB-BSSs.

\section{OBSERVATIONS AND DATA REDUCTION}

\subsection{The data sets}
\label{sec:data}
The present work is based on a combination of different
high-resolution and wide-field data-sets. The high resolution set consists of a
series of WFPC2 and ACS images
taken at various wavelengths ranging from the UV to the optical
bands. The WFPC2 images (Prop 8709, P.I. Ferraro) were obtained through the UV filters $F160BW$
and $F255W$ with total exposure times
$t_{\rm exp}= 1800$\,s and $t_{\rm exp}= 2000$\,s respectively, and through the
optical filters $F336W$ and $F555W$ with exposure times $t_{\rm exp}= 1800$\,s
and $t_{\rm exp}=106$\,s.  The center of the cluster is located in the WF2
chip (pixel scale $\sim 0.1\farcs {\rm pixel}^{-1}$). The photometric
reduction of these data was performed using ROMAFOT (Buonanno et
al. 1983) a package developed to obtain accurate photometry in crowded
regions and specifically optimized to handle under-sampled point
spread functions (Buonanno \& Iannicola 1989). 
The ACS data-set is a series of images in $F606W$
($\sim V$) and $F814W$ ($\sim I$) with $t_{\rm exp}=20$\,s and $t_{\rm
exp}=20$\,s (Prop. 10775, P.I. Sarajedini). The images were corrected
for geometrical distortions and effective flux (Sirianni et al. 2005).
The photometric reduction was performed using the photometric package
SExtractor (Bertin \& Arnouts 1996).  The wide field set is composed
of data obtained with 3 different instruments:

\begin{description}
\item[a)] EMMI-ESO-NTT -- $B$ and $V$ images (with $t_{\rm exp}=40$\,s
and $t_{\rm exp}= 20$\,s) were taken with the ESO Multi Mode Instrument
(EMMI) at the NTT during an observing run in July 2007
(P.I. Ferraro, Prop 079.D-0325).  We used the EMMI Red CCD that is
composed of 2 chips of $2048\times4093$ pixels each with a pixel scale of
about $0.33\farcs {\rm pixel}^{-1}$ and an effective field of view (FOV) of
about $9.0\farcm  \times 9.9\farcm $.  The images were corrected for
bias and flat field by using standard IRAF tools. The data reduction
was performed with SExtractor (Bertin \& Arnouts 1996).

\item[b)] MEGACAM-CFHT -- A combination of short and long MEGACAM
exposures taken through the $g$ ($t_{\rm exp}=24$\,s and $t_{\rm
exp}=240$\,s) and $r$ ($t_{\rm exp}=48$\,s and $t_{\rm exp}=480$\,s)
filters was retrieved from the Canadian Astronomy Data Centre
(CADC4). The wide field imager MEGACAM is mounted at
Canadian-French-Hawaiian Telescope (CFHT) and consists of 36 CCDs of
$2048 \times 4612$ pixels each.  For this work we used two different
pointings in which the cluster center is located between chip \#27 and
chip \#36, and \#19 and \#28 respectively. This allowed a coverage of
an area of $2\times1\,{\rm deg}^{2}$ and a complete sampling of the cluster
well beyond its tidal radius.  The data were pre-processed,
astrometrized and calibrated by using the Elixir pipeline. We performed the
data reduction using SExtractor (Bertin \& Arnouts 1996).  Each chip
in each image was reduced separately and then combined with all
the others for obtaining a catalog with $g$ and $r$ magnitudes and positions of
the detected stars. 

\item[c)] GALEX -- A complete coverage of the cluster in the UV bands
was obtained using GALEX data (FOV of about 1\,deg$^{2}$) through the
$FUV$ (1350--1750\AA) and $NUV$ (1750--2800\AA) detectors (program
GI-056, P.I. Schiavon).  Because of the high concentration of M2 and
the low angular resolution of the GALEX channels ($4\arcsec$ in $FUV$
and $6\arcsec$ in $NUV$) we used the GALEX data only for $r\geq
200\arcsec$ from the center of gravity (see below).  The reduction of GALEX
data was performed independently for each filter with
DAOPHOTII/ALLFRAME (Stetson 1987).
\end{description}

\section{Definition of the photometric catalogs} 
\subsection{Astrometry and photometric calibration}

All the catalogs were put on the absolute astrometric system using a
large number of stars in common with the Sloan Digital Sky Survey
(SDSS) catalog. As a first step we obtained the astrometric solution
of the 72 chips of MEGACAM by using the procedure described in Ferraro
et al. (2001, 2003) and a specific cross-correlation tool.  All the
stars in common with the GALEX, EMMI and HST samples were then used as
secondary astrometric standards in order to put all the catalogs in
the same astrometric system.  Several hundred astrometric
standards have been found in each step, allowing a very precise
astrometry for each catalog. At the end of the procedure the estimated
error in the absolute positions, both in right ascension ($\alpha$)
and declination ($\delta$) is about $0.2\farcs$

All the WFPC2 magnitudes ($m_{160}$, $m_{255}$, $m_{336}$ and
$m_{555}$) were calibrated in the STMAG system using the equations and
zeropoints listed in Holtzmann et al. (1995) and the same procedure
described in Ferraro et al. (1997, 2001).  Then the stars in common
between the other catalogs and the WFPC2 sample were used to transform
all the magnitudes to the same photometric system. In particular, the
$F606W$ of the ACS catalog, the EMMI instrumental $V$ magnitudes and
MEGACAM $g$ magnitudes were transformed to the $V$ STMAG by using
appropriate color equations. The EMMI B instrumental magnitudes were
put in the STMAG system. The ACS $F814W$ magnitudes were calibrated in the
STMAG system using the prescriptions of Sirianni et al. (2005), and
the $r$ MEGACAM mag was transformed to the SDSS system. The GALEX
instrumental $FUV$ and $NUV$ magnitudes were calibrated to STMAG
system using the stars in common with the WFPC2.

\subsection{Center of Gravity}

The center of gravity has been obtained following the procedure
adopted in our previous work (see for example Lanzoni et
al. 2007b). A first estimate of the cluster center was performed by
eye on the WF2 chip of the WFPC2 image, then the exact measure of
$C_{\rm grav}$ was obtained by means of an iterative procedure that
averages the absolute positions of stars lying within $\sim10\arcsec$
from the first guess center. In order to avoid biases and spurious
effects, we considered two samples with two different limiting
magnitudes ($V<19.7$ and $V<19.2$). The values of $C_{\rm
grav}$ obtained with the two samples agree within
$1\arcsec$. We adopt the mean value as the best estimate of $C_{\rm
grav}$: $\alpha = 21^{\rm h} 33^{\rm m} 27^{\rm s}~(RA= 323.3623340)$ and $\delta =
-0^{\circ} 49\arcmin 22.8\farcs ~ (Dec= -0.82304665)$ .
This new determination is substantially different from the
center reported by Harris et al. (1996) on the basis of the surface
brightness profile and using photographic plates: our $C_{\rm grav}$
is located at $\sim 35\arcsec$ west ($\Delta\alpha \sim
35\arcsec $ , $\Delta\delta \sim 0\arcsec$) from Harris center.

\subsection{Sample definition}

Once all the data-sets have been photometrically homogenized and put
in the same reference frame, and the cluster center has been
determined, we have built a single catalog by combining the following
sub-samples: \emph{i)} the WFPC2 sample, composed of all the stars
detected in the WFPC2 FOV; \emph{ii)} the ACS sample, comprising all
the stars in the ACS FOV complementary to the WFPC2 one; \emph{iii)}
the EMMI sample, complementary to the previous two and including only
stars with distance $r<200\arcsec$ from $C_{\rm grav}$ and \emph{iv)}
the MEGACAM/GALEX sample made of stars with $r\geq 200\arcsec$ included
in the MEGACAM FOV (of course only a fraction of these stars also has
GALEX magnitudes). The criteria used for these definitions have
been chosen to sample the highly crowded central
regions of the cluster with the highest spatial resolution and UV band
data (thus to maximally limit the effects of photometric errors and
stellar blends), while covering the entire cluster extension by means
of wide-field images. The maps of the adopted samples are shown in
Figures 1 and 2. In Fig.~3 the ($V,~U-V$) CMD of the WFPC2 sample is shown.

\subsection{Density profile} 
 
We have determined the projected density profile of M2 by measuring
the star counts over the entire cluster extension. Only stars with
$15.2<V<19.2$ in the combined sample, covering the cluster extension 
from $C_{\rm grav}$ to $r=1800\arcsec$ were considered (see Figs. 4 and 5). The area was divided in 36 annuli all
centered on $C_{\rm grav}$.  Each annulus was divided into an
adequate number of sub-sectors in which the stellar density has been
calculated as the ratio between the number of stars and the sub-sector
area. For each annulus the resulting density is given by the average
of the corresponding sub-sector densities and the error is quoted as
the square root of the variance of the sub-sector densities. In this
procedure we have also taken into account the incomplete area coverage
of the most external annuli and the largest CCD gap in the MEGACAM
FOV.

The observed density profile is plotted in Fig.~6. The sample nicely
covers the entire cluster extension. The four outermost annuli (with
$r>600\arcsec$) show a flattening of star counts giving a direct
estimate of the stellar background in the cluster direction: for
$15.2<V<19.2$ the background star density is $\sim0.7$~stars/arcmin$^{2}$. The
observed profile is well reproduced by an isotropic single-mass King
model with concentration $c\simeq1.51$ and core radius
$r_{c}\simeq17\arcsec$. The corresponding tidal radius is
$r_{t}\simeq550\arcsec$. Since there is an uncertainty of about 15$\%$
in the determination of $r_{t}$, in our analysis below we will consider
all stars lying within $r<650\arcsec$.  The newly determined cluster parameters 
are substantially
different from those reported by Harris et al. (1996) based on the luminosity 
center and the surface brightness 
distribution ($c=1.8$ and $r_{c}=20\arcsec$) and from the even
higher concentration model found by Pryor \& Meylan (1993; $c=1.9$ and
$r_{c}=20\arcsec$).  As shown in Fig.~6 (dashed line), a King model
with the parameters quoted by Harris et al.  (1996) does not reproduce
the observed profile.  On the contrary, a reasonable agreement (within
the errors) is found with the values estimated by McLaughlin \& van
der Marel (2005; $c=1.59$ and $r_{c}=19\arcsec$). Assuming a distance
modulus $(m-M)_V=15.49$ and a reddening $E(B-V)=0.06$ (Harris et
al. 1996) we find a real distance $d\simeq 12.5$\,kpc, and a core radius
$r_{c}\simeq 1.02$\,pc.

The best-fit model reproduces the observed profile out to $400\arcsec$
very well, while at larger distances the observed star counts show an
excess with respect to the model. While this discrepancy is not
statistically significant, it deserves further investigation since it
could be the signature of tidal distortion in the outer regions (see
Leon et al. 2000 for more details). Another interesting feature of
density profile is that the innermost point seems to deviate from the
canonical flat-core King model. This is also worthy of future
investigation since similar features might be related to the presence
of an intermediate mass-black hole (e.g. Miocchi 2007, Lanzoni et
al. 2007c).

\section{THE BSS AND REFERENCE POPULATION SELECTION}
\label{sec:BSS}
\subsection{The BSS selection}

In this section we describe the procedure that we have followed to
select the BSS population and to construct the BSS radial
distribution in M2.   At the UV wavelengths, hot populations like BSSs and
extreme-HB stars are the brightest objects, while cool populations
(like red giant branch -- RGB -- stars) appear quite faint (see Figs. 7 and 10).  Because of
this, we always prefer to use the UV-CMD as the reference plane for the
BSS selection. Moreover, since the HST spatial
resolution dramatically reduces problems connected with crowding and
blends, we have primarily selected the BSS population by considering the WFPC2 sample in the 
($m_{255}$, $m_{255}-U$) plane. In order to avoid contamination from the sub-giant branch
(SGB) stars, we selected only stars with $m_{255}<19.55$, that is about
1 magnitude brighter than the turn-off (TO) point
($m_{255}\simeq 20.5$). The number of BSS thus selected in the WFPC2 sample
is 82.

As in previous studies, we used the UV-selected BSS in common with the
ACS sample to define a selection box in the ($V$, $V-I$) plane. We
have adopted a limiting magnitude $V\sim19.2$, and the red edge is at
$(V-I)=0.55$ (see Fig.~8). The total number of BSS found in the ACS
sample is 20. In the EMMI catalog the BSS have been selected in the
($V$, $B-V$) CMD, using the same cut in the $V$ filter as for ACS
sample. Considering the quality of the diagram the color limit was set
to $(B-V)<0.32$ to avoid spurious detections and blends from TO and
SGB stars: 9 BSS have been selected in this way (see Fig.~9). In
the most external region sampled by our observations ($r\geq
200\arcsec$) the combination of the MEGACAM and the GALEX samples
allows the construction of an UV CMD. Since both the GALEX $NUV$ and
the HST $m_{255}$ magnitudes have been calibrated on the STMAG
photometric system (see Sect. 3.1), we have used the same threshold
($NUV<19.55$) adopted for the WFPC2 sample to define the selection box
in the ($NUV$, $NUV-V$) plane. The result is shown in Fig.~10, where
12 BSS have been selected for $r\geq200\arcsec$. The right panel of
Fig.~10 shows the location of the selected BSS in the ($V$,
$V-r$) plane.  In summary a total of 123 BSS have been selected in M2
(see Table 1).

\subsection{The reference populations}

As discussed in other papers (see Ferraro 2006a and references in
Dalessandro et al. 2008a) we also need to select a reference
population which is representative of the ``normal'' cluster
population. As in other works of this series, we have used the HB and
RGB stars as reference populations. The selection of the RGB stars has
been performed in the optical planes. For all of the samples a
magnitude cut at $V<18$ has been adopted. However for our analysis
only stars with $V>16$ were used in order to avoid saturated stars in
the ACS and MEGACAM/GALEX sample (Fig.~8 and Fig.~10).  The color
limits of the selection boxes have been chosen to follow the RGB ridge
mean line in each CMD while avoiding regions with high probability of
field star contamination (the selected RGB stars are marked with empty
squares in Fig.~8, 9 and 10).  We found 2121 RGB within $r<650\arcsec$
(1223 in WFPC2, 460 in ACS, 270 in EMMI and 168 in MEGACAM/GALEX
samples, respectively). The magnitude range of the RGB reference
population is the same as that adopted for the ''faint'' RGB discussed
below. 

In the WFPC2 and MEGACAM/GALEX samples the HB stars have been selected
on the basis of their positions in the ($m_{255}$, $m_{255}-V$) and (NUV,NUV-V) CMDs respectively
(see left panel of Fig.~10 for the wide-field sample). The positions in the optical 
MEGACAM/GALEX plane of the selected HB stars 
(Fig.~10 right panel) have been used to
define the selection box for the ACS and EMMI samples (see Figs.~8 and
9).  By cross-correlating our catalog with the catalogs of RR~Lyrae
stars found by Lee \& Carney (1999) and Lazaro et al. (2006), we have
identified all of the 42 known variables (they are marked as asterisks
in Fig.~7, 8, 9 and 10) and we have included them in our HB sample. The
total number of HB stars within $r<650\arcsec$ is 875 (525 in WFPC2,
184 in ACS, 104 in EMMI and 62 in MEGACAM/GALEX samples).

\section{Results}
\subsection{The BSS radial distribution}
 
Having defined the reference populations we can now examine the BSS
radial distribution. The BSS cumulative radial distribution is shown
in Fig.~11 with the distributions of the HB and RGB stars shown for
comparison. The BSS population is more segregated in the central
regions and less concentrated in the outer parts than either the HB
and the RGB stars. The KS test gives a probability of $\sim 10^{-6}$
($4\sigma$ significance level) that the radial distribution of the BSS
is extracted from the same parent distribution of the reference
population.

For a more quantitative analysis we computed the population ratios
$N_{\rm BSS}/N_{\rm HB}$ and $N_{\rm BSS}/N_{\rm RGB}$ (where $N_{\rm
pop}$ is the number of stars belonging to a given population) in 6
concentric annuli centered on $C_{\rm grav}$.  To do this we had to
evaluate the impact of field star contamination on each
population. The field stars predominantly lie in a vertical sequence
at $0.2<(V-r)<0.5$ and dramatically affect the RGB population (see Fig~5 right panel). An
estimate of the field star contamination can be directly obtained from
our sample by considering an annulus at $1900\arcsec<r<2400\arcsec$
($\sim70\%$ of which is sampled by the MEGACAM data) far beyond the
tidal radius of the cluster ($r_{t}\sim550\arcsec$). We
counted the number of field stars in this annulus lying within the
BSS, HB and RGB selection boxes shown in Figs.~8, 9, and 10, and we
derived the following values for their density: $\rho_{\rm
BSS}\sim0.01$~stars/arcmin$^{2}$, $\rho_{\rm
RGB}\sim0.06$~stars/arcmin$^{2}$, while no field stars have been found
within the HB selection box. These values have been used to
statistically decontaminate the star counts in each annulus.
 
The star counts for each annulus are listed in Table~2. These values
have been used to compute the ratios $N_{\rm BSS}/N_{\rm HB}$ and
$N_{\rm BSS}/N_{\rm RGB}$. The radial distribution of these ratios is
shown in Fig.~12 (central and upper panels, respectively). They are
clearly bimodal, with a high BSS frequency in the central and outer
regions, and with a broad minimum at about 120$\arcsec$
($\sim9r_{\rm c}$) from $C_{\rm grav}$. On the contrary the $N_{\rm
HB}/N_{\rm RGB}$ ratio (plotted in the bottom panel of Fig.~12) shows a
flat distribution across the cluster extension, as expected for
``normal'' populations. As a further confirmation of the BSS
bimodality, we also computed the double normalized ratio 
as defined in Ferraro et al. (1993):

\begin{displaymath}							   
R_{\rm pop}=\frac{N_{\rm pop}/ N^{\rm tot}_{\rm pop}}{L^{\rm samp}/ L^{\rm samp}_{\rm tot}}.
\label{eq:Rpop}
\end{displaymath}

\noindent where pop = BSS, HB. The total sampled luminosity ($L_{\rm
  tot}^{\rm samp}$), as well as the luminosity sampled in each annulus
  ($L^{\rm samp}$), has been estimated from the King model by using the cluster structural
  parameters, distance modulus and reddening quoted in Section 3.3,
  and the central surface brightness reported by Harris et
  al. (1996). The incomplete spatial coverage due to the largest
  ($\sim 1\arcmin$) gap between the MEGACAM CCDs has been taken into
  account.  As shown in Fig.~13, $R_{\rm HB}$ is constant with a value
  close to 1 out to $r=650\arcsec$. This is just as expected: the fraction of HB (as any post-MS) stars is proportional to
the fraction of sampled light, as shown in Renzini \& Fusi Pecci
(1988). Conversely the radial distribution of
  the BSS double normalized ratio ($R_{\rm BSS}$) confirms the bimodal
  behaviour: it is peaked in the central regions, decreases to a
  minimum value at about $9r_{c}$ and then rises again in the cluster
  outskirts.

The location of this minimum at $r\sim9r_{c}$ can be related to the
dynamical evolution of the cluster and in particular to the radius of
avoidance ($r_{\rm avoid}$).  This parameter is defined as the radius
within which all the stars as massive as $1.2 M_{\odot}$ (the assumed
mass for BSSs) have already sunk to the center because of mass
segregation (Mapelli et al. 2004, 2006). Using the dynamical friction
time-scale formula (e.g. Mapelli et al. 2006) under the assumption of
a cluster age $t=12$\,Gyr, a central velocity dispersion of
$\sigma_{0}=8.2\,{\rm km\,s^{-1}}$ (Pryor \& Meylan 1993), we obtained $r_{\rm
  avoid}\sim 7r_{c}$. This position is fully compatible with the position of
the observed minimum.  

\subsection{The AGB problem}

Beccari et al. (2006a) found a significant overabundance of
AGB stars in the very central regions of 47~Tuc. This excess could be due to contamination of 
genuine AGBs by massive
(1.1--$1.5\,M_{\odot}$) objects in late evolutionary stages (e.g. in the horizontal branch
phase, as suggested by Sills et al. 2008). Presumably these objects arise from binary systems (mainly
BSSs) segregated in the cluster core because of dynamical effects.  To
search for a similar result in M2, we used the WFPC2 and the EMMI
sample where the brightest evolutionary sequences are well defined up
to the RGB tip at $V\sim13$.  We selected AGB stars in the ($V$, $U-V$) plane for the WFPC2 sample and in the 
($V$, $B-V$) for the EMMI
sample as shown in Fig.~14. It was not possible to use either the ACS
or the MEGACAM/GALEX samples because of saturation problems.

To study the radial distribution we divided the covered region into 5
concentric annuli centered on $C_{\rm grav}$ and counted the number of
AGBs and HBs lying in each annulus. It was not possible to do a
statistical decontamination of the AGB population because the MEGACAM/GALEX sample saturates at
$V\sim15.5$. However, we would expect that in the central regions it does not appreciably affect the
observed radial distribution. Fig.~15 upper panel shows the behaviour of the population ratios $N_{\rm
AGB}/N_{\rm HB}$  as a function of the distance from the cluster center. As apparent from the figure,
while the mean value of the 4 outermost annuli is $\sim 0.12\pm0.03$, fully consistent with the value
expected from the evolutionary timescales (Renzini \& Fusi Pecci 1988), the ratio turns out to be higher
($\sim 0.19\pm0.03$) in the outermost annulus (corresponding to $r_{c}$). This central
overconcentration of the AGB population corresponds to an excess of about
$30\%$ (or 9-10 more stars) in the first annulus. This value is compatible with the 
life-times and populations ratios computed by Sills et al. (2008) for evolved collisional products,
supporting the idea of a possible contamination by evolved BSS. To further
investigate this feature we also computed the double normalized ratio.
The incomplete spatial coverage has been taken into account.  
The radial distribution of $R_{\rm AGB}$ (see Fig. 15 bottom panel) fully confirms this behaviour,
showing a central peak ($R_{\rm AGB}\sim1.4$) within $r_{c}$, while in the outer part the ratio remains
constant at $R_{\rm AGB}\sim 1$ fully in agreement with $R_{\rm HB}$.

Purely on the basis of small number statistics introduced by binning,
 the AGB central peak is marginally significant ($< 2\sigma$). However
 the significance of the peak can also be evaluated with a KS test on
 the cumulative distribution, which is shown in
 Fig.~\ref{agb-dist}. The probability that the AGBs are drawn from a
 different distribution from the HBs is 93\% ($\sim 1.8\sigma$). The BSS distribution is
 also shown in Fig.~\ref{agb-dist}. While AGBs are more
 concentrated than HBs, they are less concentrated than BSSs, with 
 a 98\% probability that they are extracted from a
 different parent family. In this respect they are different from the
 AGBs in 47~Tuc where AGBs and BSSs have similar radial
 distributions.

\subsection{Color gradients}

Sohn et al. (1996), hereafter S96, found that M2 has a radial color gradient, in the
sense that the central regions are bluer than the outer parts, with a
variation of about $(B-V)\sim0.1$. To investigate this interesting
feature we computed the ($U-V$) integrated color within
$90\arcsec$ from $C_{\rm grav}$ which approximately corresponds to the region used by
S96. 
We divided the WFPC2 sample in 5 concentric
annuli (the first corresponding to $r_{c}$), and computed the 
color of each annulus from the resolved stars by considering three different magnitude cuts:
$V<16$, $16\leq V<20$ and $V<20$. As shown in Fig.~17 (upper panel) we found
that when only the brightest stars are included ($V<16$, black and
open dots in Fig.~17) a color difference $\Delta(U-V)\sim0.18$ between
the center (bluer) and the outer annuli is apparent. Even if this is a less than $2 \sigma$ 
result, it is consistent with the finding of S96.
When also fainter stars are included (i.e. for $V<20$), the color gradient decreases, and if the
brightest stars are excluded ($16\leq V<20$) it completely disappears and ($U-V$) remains constant
all over the considered radial range.
To further investigate this behaviour we made the same computation for the
ACS sample using the ($V-I$) color. In this sample saturation occurs
at about $V=15$, so the test is limited to the population with
$16<V<20$.  No color gradient is visible in the bottom panel of
Fig.~17.
Our results therefore indicate that the observed color gradient is due to the brightest stars and
not to an over-concentration of BSSs or blue faint objects. This seems in disagreement with the
conclusion of S96,
who found the color gradient only when using resolved stars with $V<16$. However, as already
discussed by these authors, the poor seeing conditions and the
spatial resolution of the instrument (0.56\farcs$\,{\rm pixel}^{-1}$)
used in their analysis did not allow them to sample all the
populations with acceptable photometric accuracy.  
To more deeply understand the origin of the detected color gradient, we further investigated the
properties of the brightest populations in the very central regions of M2. Since the AGB is
0.2-0.3 mag bluer than the RGB in ($U-V$), we first investigated whether the AGB central excess 
(Sect.5.2) could account for the observed color gradient. We therefore artificially cancelled the
AGB central peak, by randomly excluding 10 stars from the innermost bin, and re-computed the
central color: this still yields a center bluer than the exterior.
Very bright RGB stars therefore
remain the only candidates. In order to test this hypothesis
we compared the radial distribution of the brightest portion of the RGB
($V<16$) in the WFPC2 sample (see Fig.~14, left panel) to the faint ($V\ge16$) one.
The radial distributions of these populations clearly show that the
brightest giants are less concentrated than the faintest ones, with a $99\%$ probability (about $2.5\sigma$) 
that they are extracted from a different parent family (see Fig.~16 and the upper panel of
Fig.~18). We have therefore re-computed the central color after having artificially increased the
number of bright RGBs in the innermost bin, thus to flatten the radial distribution of the
bright-to-faint RGB ratio (to this purpose, we have randomly extracted 25 bright RGBs from the
observed luminosity function). This completely removes the color gradient (bottom panel of Fig.~18). 
Hence we conclude the the
color gradient found by S96 and confirmed here is due
to a deficit of bright RGB stars in the center rather than a
surplus of fainter blue stars.

\section{Summary}

The BSS population of M2 can be characterized as what is emerging as
{\em ''normal''}: a bimodal radial distribution with a minimum in the zone of
avoidance, and with a value of the central BSS specific frequency 
($N_{\rm BSS}/N_{\rm HB}$) which is also typical. Bimodal distributions are a
very common feature of the Galactic GC BSS populations (Dalessandro et
al. 2008a).  Only two clusters, NGC~2419
and $\omega$\,Cen, deviate significantly from this pattern. Both of
these systems are very large. There is even some doubt that
$\omega$\,Cen is a true GC (Bekki \& Freeman 2003). Of
the bimodal clusters only two, NGC~6388 (Dalessandro et al. 2008a) and
NGC~5024 (Beccari et al. 2008), have minima in their BSS radial
distributions which differ significantly from $r_{\rm
  avoid}$. Presumably this arises because of a lower efficiency of the
dynamical friction in these two clusters, for reasons yet to be
explained.
 
As Beccari et al. (2006) found for 47~Tuc, we find an excess of AGB
stars in the center of M2. Because of the smallish sample size, the
excess is only marginally significant, and unlike in 47~Tuc, the AGB population is not
as concentrated as the BSS one. 

In agreement with S96 we find that the integrated color of the central
region of M2 is bluer that the exterior. We show that this color
gradient is due to a deficit of bright RGB stars, and not to an excess
of faint blue objects, such as BSS or HB stars.  A similar deficit of
bright RGB stars has also been found in the very massive GC NGC~2808
(Sandquist et al. 2007). They do not explore the radial dependence of
their result, and neither of the two mechanisms they discuss for
producing a deficit (neutrino losses and extra mass loss) would have
an obvious radial dependence. We view our AGB surplus and bright RGB
deficit as suggestive---given the short lifetime in these phases it is
impossible to do better than $2\sigma$ in M2 or any single cluster. If
similar results are found in other clusters, there would be
interesting consequences for stellar evolution theory and stellar
population studies. Given this, it would be highly desirable that
future photometric studies of GCs were designed in such a way that
unsaturated photometry of the brightest stars was possible.

\acknowledgements 
This research was supported by  the {\it Progetti Strategici di Ateneo 2006} 
granted by the University of
Bologna. We also acknowledge the financial support of the Agenzia Spaziale Italiana 
(under the contract ASI-INAF I/016/07/0),
and the Ministero dell'Istruzione, dell' Universit\'a e della Ricerca. 
ED is supported by INAF - Osservatorio Astronomico di Bologna, he 
acknowledges the {\it Marco Polo} programme of the Bologna
University for a grant support, and the Department of Astronomy of the 
University of Virginia for the hospitality during his stay, when
part of this work was done. RTR is partially supported
by STScI grant GO-10845. RTR \& RS are partially supported by GALEX
grants GI1-56 \& GI4-33.  
This research has made use of the ESO/ST-ECF
Science Archive facility which is a joint collaboration of the European
Southern Observatory and the Space Telescope - European Coordinating
Facility.

%--Table 1
\begin{deluxetable}{lcccccccc}
\footnotesize
\tablewidth{15.5cm}
\tablecaption{The BSS population of M2}
\startdata \\
\hline \hline
Name    &    RA       &   DEC       &  m$_{255}$ &  U   &  B  &  V  &  I  &  r  \\
        & [degree]    &  [degree]   &            &      &     &     &     &     \\
\hline
BSS   1 & 323.3714411 & -0.8178864 & 18.296 & 18.526  & 0.000 & 17.276 & 15.678 &  --- \\ 
BSS   2 & 323.3696276 & -0.8177717 & 17.413 & 17.460  & 0.000 & 17.828 & 16.353 &  --- \\ 
BSS   3 & 323.3634359 & -0.8316575 & 18.095 & 17.625  & 0.000 & 17.147 & 16.653 &  --- \\ 
BSS   4 & 323.3622994 & -0.8218842 & 18.652 & 18.077  & 0.000 & 17.638 & 16.720 &  --- \\ 
..........  & & & & & \\
\hline
\enddata
\tablecomments{The complete table is available in electronic form.}
\label{tab:bss}
\end{deluxetable}

%---table 2
\begin{deluxetable}{rrlrlc}
\tablecolumns{6}
\footnotesize
\tablecaption{Number Counts of BSS, HB, and RGB Stars, and Fraction of
Sampled Luminosity}
%\tabletypesize{\scriptsize}
%
\tablewidth{9cm}
\startdata \\
\hline \hline
 {$r_i\arcsec$} &{$r_e\arcsec$} &  {$N_{\rm BSS}$} & {$N_{\rm HB}$}  & {$N_{\rm RGB}$}   & {$L^{\rm samp}/L_{\rm tot}^{\rm samp}$}  \\
\hline
   0 &   20 & 54  & 171   &  454   & 0.20 \\
  20 &   50 & 27  & 260   &  636   & 0.30 \\
  50 &  100 & 20  & 242   &  513   & 0.25 \\
 100 &  200 & 10  & 141   &  348$\,$(2) & 0.18 \\
 200 &  300 &  7  &  40   &   94$\,$(3)  & 0.05 \\
 300 &  650 & 4$\,$(1) &  21    &   59$\,$(12) & 0.02 \\
\hline 
\enddata
\tablecomments{The values listed out of the parenthesis correspond to the
number of stars assumed to belong to the cluster (and thus used in the analysis),
while those in the parenthesis are estimated to be contaminating field
stars (see Sect. 5.1).}
\label{tab:counts}
\end{deluxetable}

\newpage
%---Fig.1 
\begin{figure}[!hp]
\begin{center}
\includegraphics[scale=0.7]{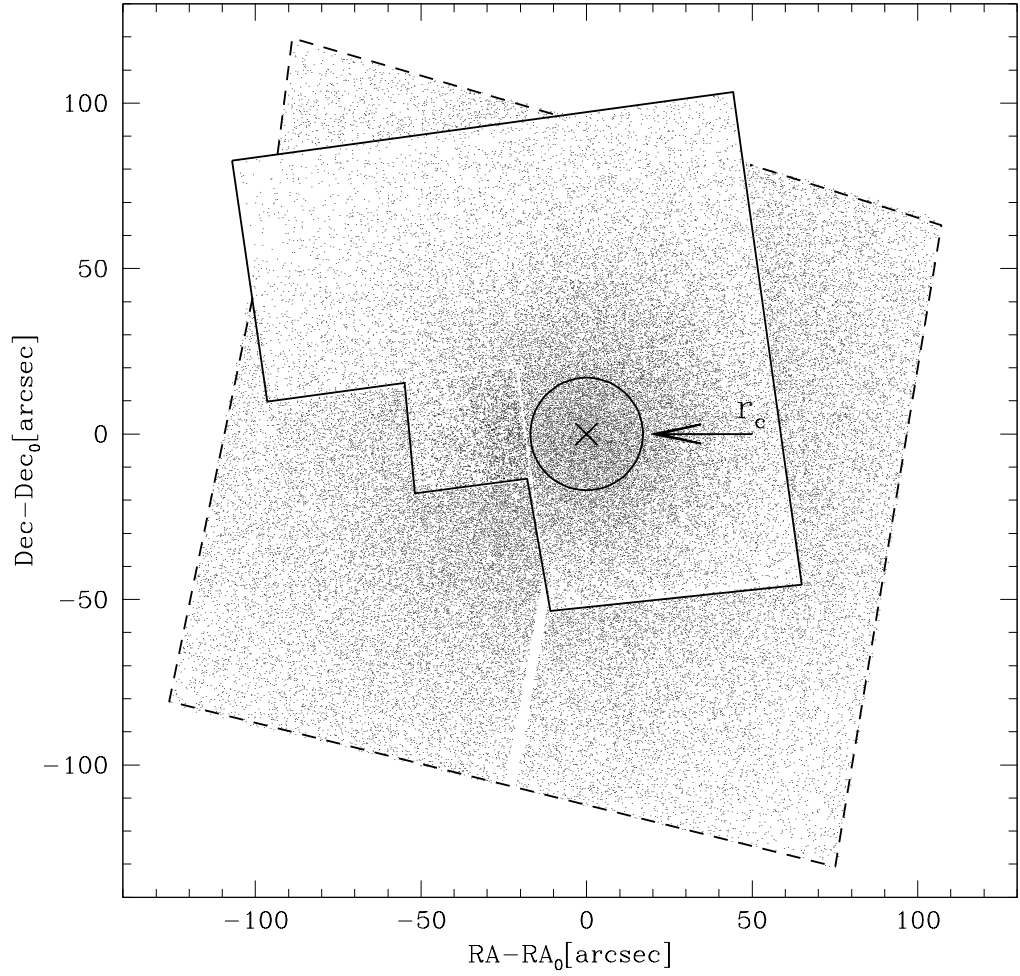}
\caption{Map of the WFPC2 sample (solid line) and the ACS sample (dashed line)
with the coordinates referred to the right ascension $RA_{0}$ and the 
declination $Dec_{0}$
of the cluster center of gravity (cross). 
The circle marks the core radius of the
cluster as determined in Sect. 3.4.}
\label{acs}
\end{center}
\end{figure}

%---Fig.2
\begin{center}
\begin{figure}[!p]
\includegraphics[scale=0.7]{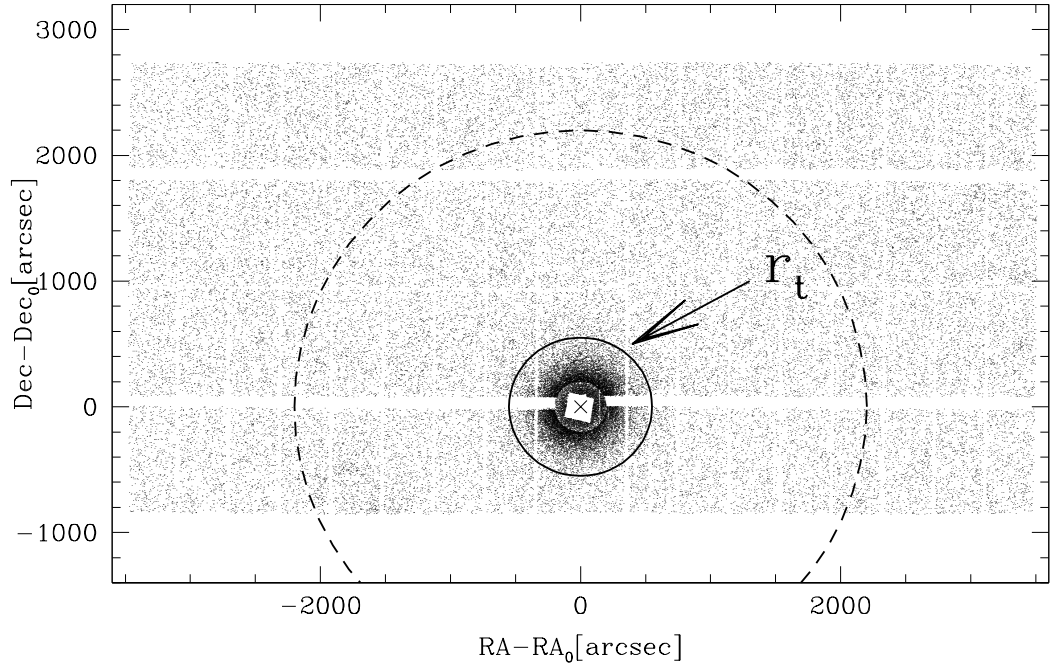}
\caption{Map of the EMMI and MEGACAM/GALEX sample. The circle with
  radius $r_{t}=550\arcsec$ marks the estimated tidal
radius, while the dashed circle indicates the GALEX FOV.}
\label{subaru}
\end{figure}
\end{center}

\begin{center}
\begin{figure}[!p]
\includegraphics[scale=0.7]{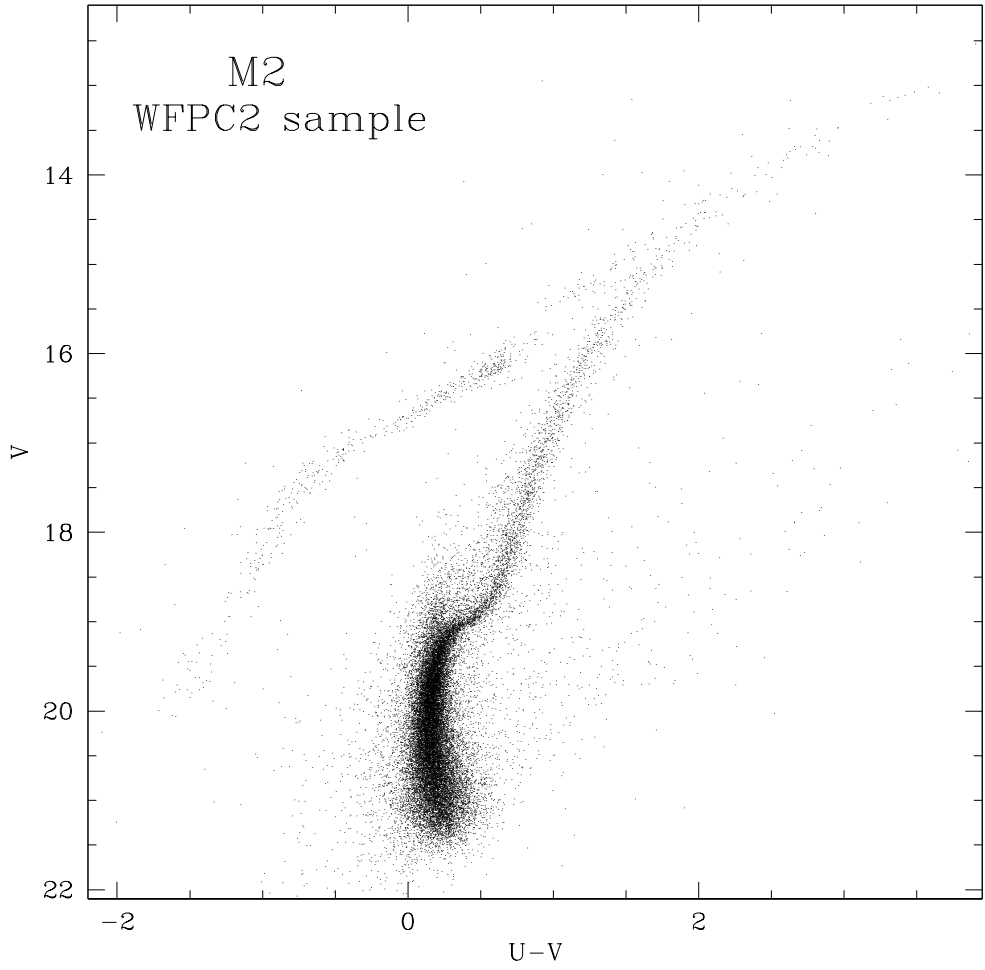}
\caption{The ($V,~U-V$) CMD of the WFPC2 sample.}
\end{figure}
\end{center}

\begin{center}
\begin{figure}[!p]
\includegraphics[scale=0.7]{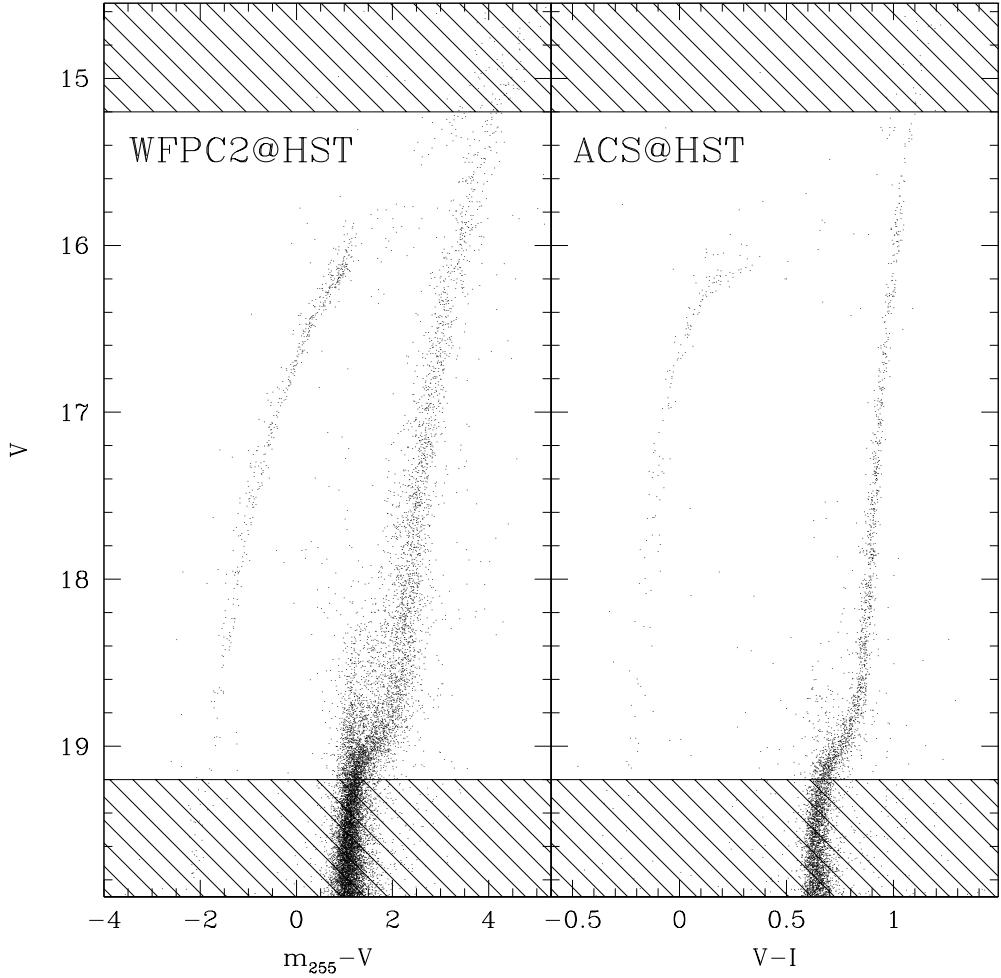}
\caption{($V$,~$m_{255}-V$) CMD of the stars lying in the WFPC2 FOV
  (left panel) and ($V$,~$V-I$) CMD of the stars detected in the ACS
  FOV complementary to WFPC2 (right panel). The shaded regions
  delimit the samples adopted to compute the star density profile (see
  Sect. 3.4).}
\end{figure}
\end{center}

\begin{center}
\begin{figure}[!p]
\includegraphics[scale=0.7]{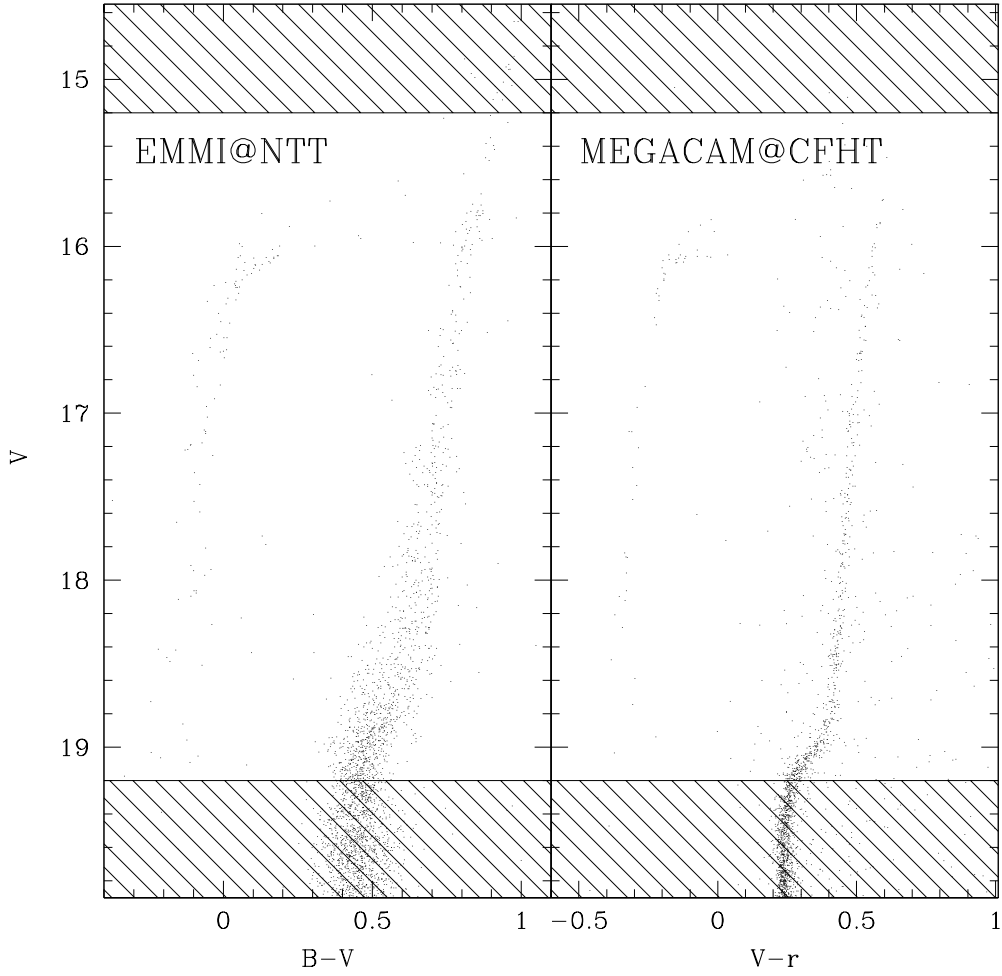}
\caption{Optical CMDs of the EMMI (left panel) and the MEGACAM sample
(right panel). The shaded region is as in Fig. 4}
\end{figure} 
\end{center}

\begin{center}
\begin{figure}[!h]
\includegraphics[scale=0.7]{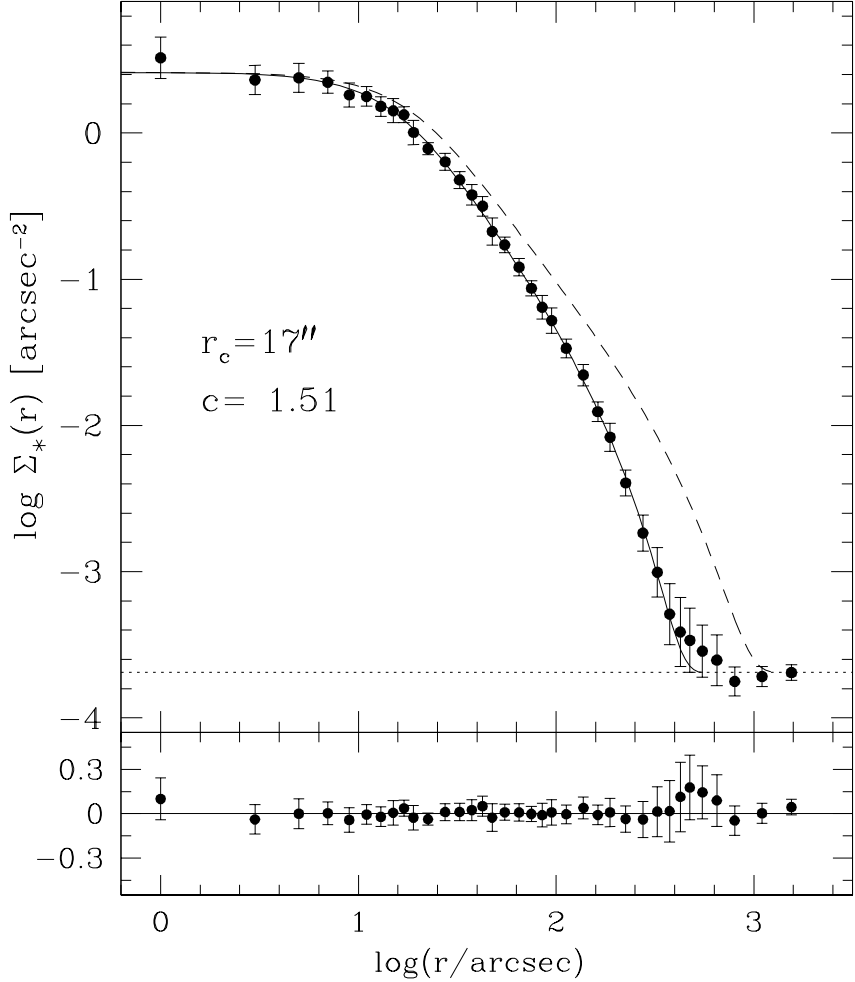}
\caption{Observed surface density profile (dots and error bars) and
  best-fit King model (solid line). The radial profile is in units of
  number of stars per square arcsecond. The dotted line indicates the
  adopted level of the background (corresponding to 0.7
 stars\,arcmin$^{-2}$ in the range $15.2<V<19.2$). The model
  parameters are $r_{c}=17\arcsec$ and $c=1.51$.  The
  lower panel shows the residuals between the observations and the
  fitted profile. The dashed line is the King-model obtained using the
  structural parameters quoted by Harris et al. (1996; see Sect. 3.4) }
\label{bbi}
\end{figure}
\end{center}

%---Fig.5
\begin{figure}[!p]
\begin{center}
\includegraphics[scale=0.7]{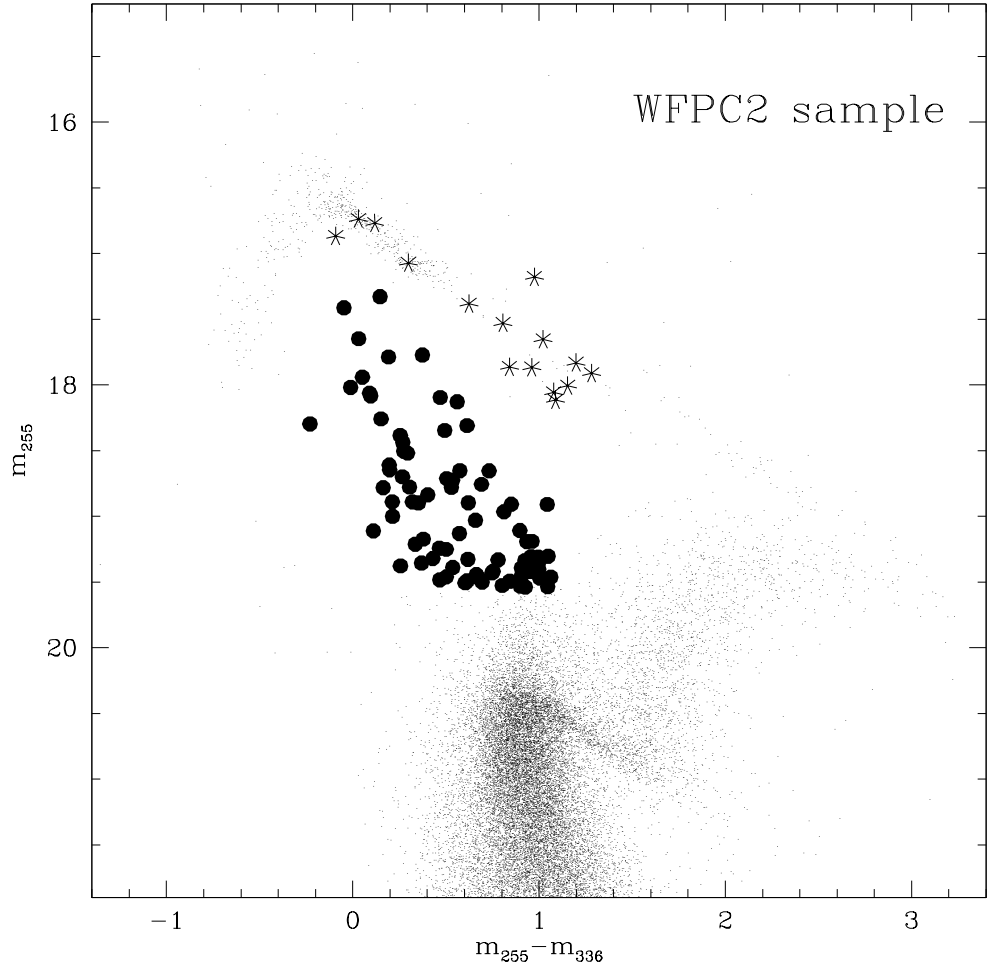}
\caption{ Ultraviolet CMD of the WFPC2 sample. The selected BSS
population is marked as filled dots, and RR Lyrae stars as asterisks.}
\label{}
\end{center}
\end{figure}

%---Fig.6
\begin{figure}[!p]
\begin{center}
\includegraphics[scale=0.7]{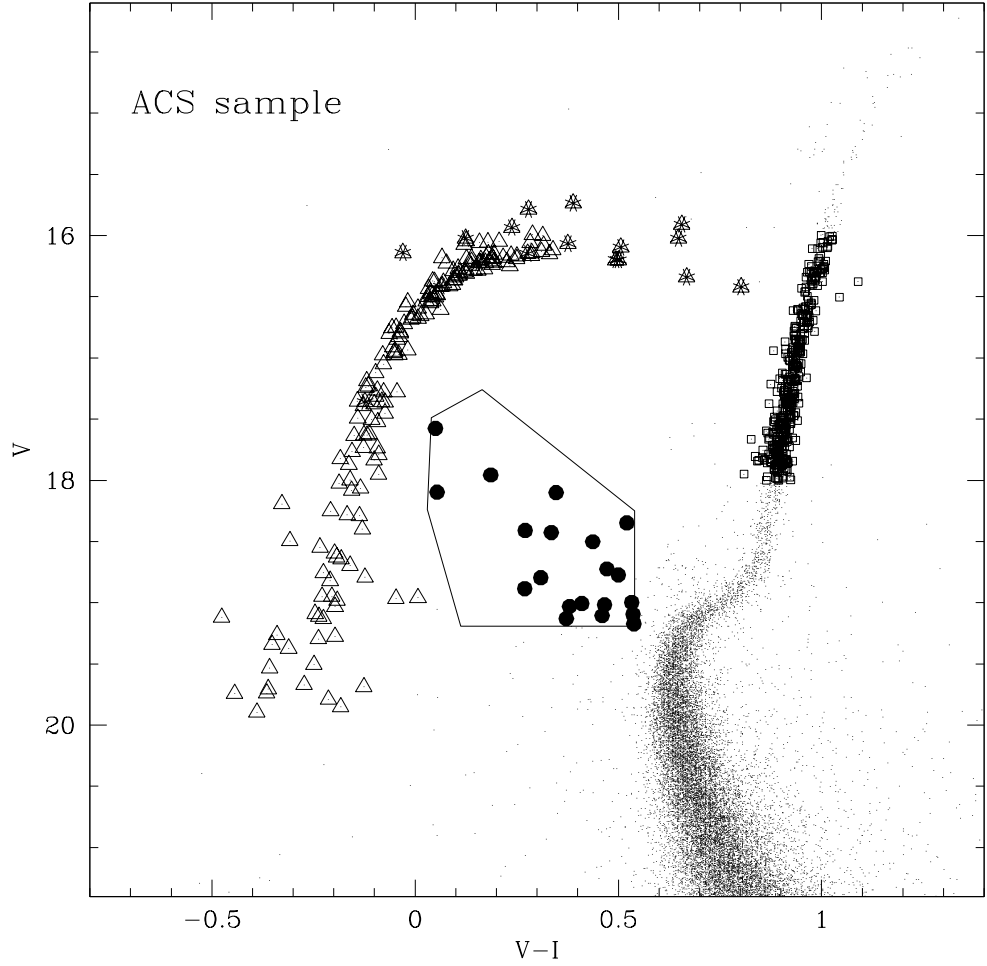}
\caption{($V$,~$V-I$) CMD of the ACS sample. The different stellar
  populations discussed in the paper are marked with different
  symbols (same as in Fig.~7 plus squares and triangles for the RGB and HB stars respectively).}
\label{vr}
\end{center}
\end{figure}

%---Fig.7
\begin{figure}[!p]
\begin{center}
\includegraphics[scale=0.7]{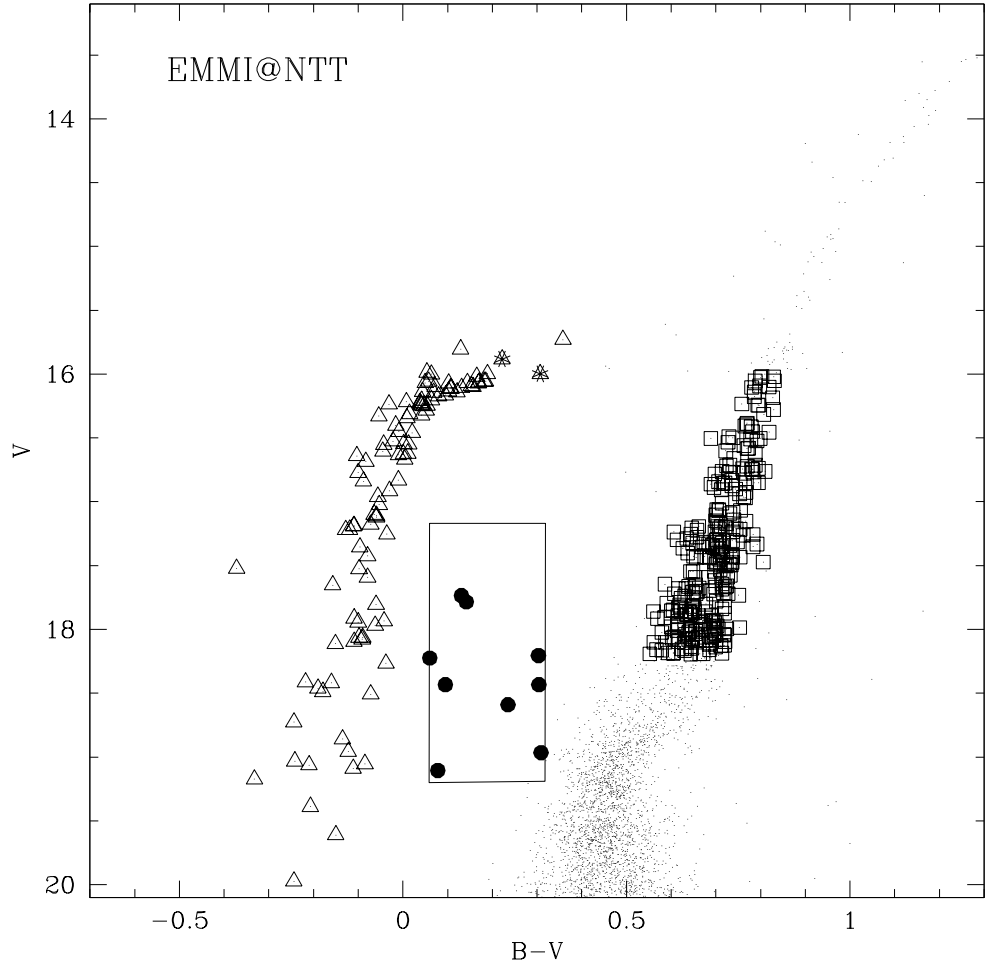}
\caption{Optical CMD of the EMMI sample. The symbols have the same
  meaning as in Fig.~8.}
\end{center}
\end{figure}

%---Fig.8
\begin{figure}[!p]
\begin{center}
\includegraphics[scale=0.7]{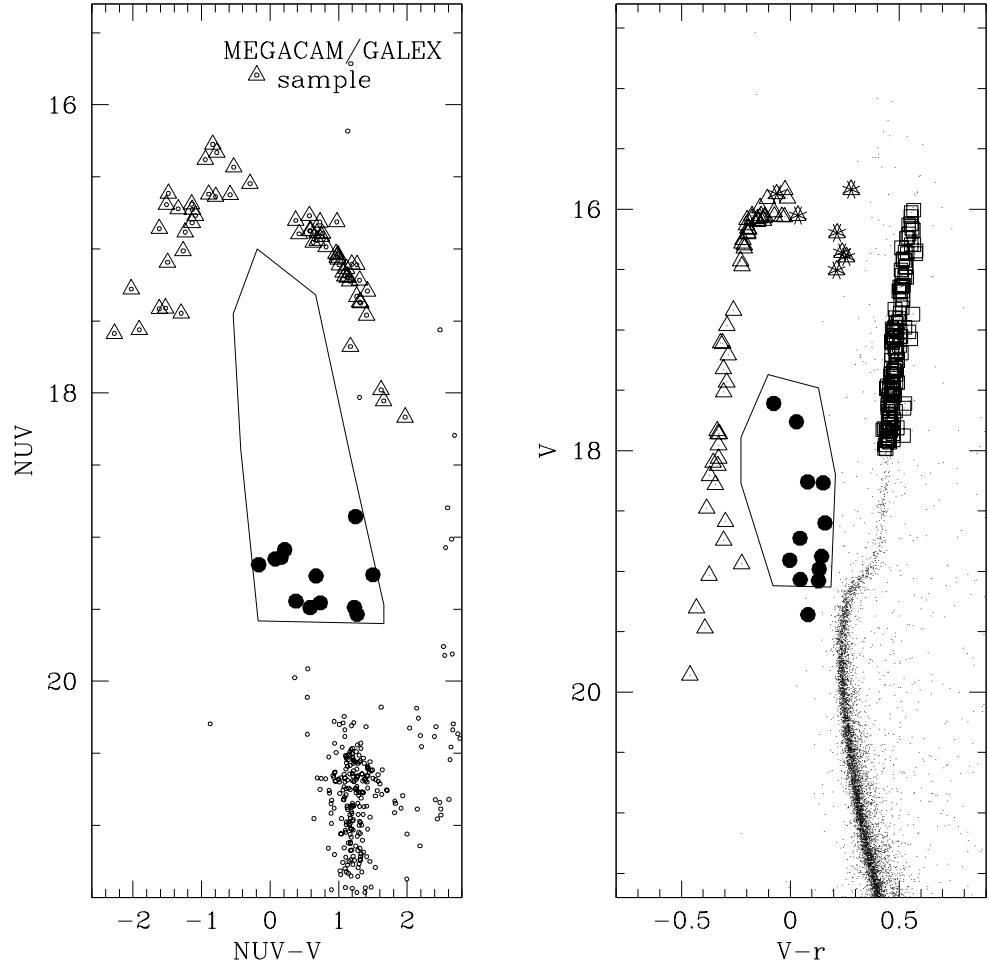}
\caption{Ultraviolet (left panel) and optical (right panel) CMDs of the MEGACAM/GALEX sample. The NUV magnitudes have
been obtained by matching the optical data with GALEX observations. The symbols have the same
  meaning as in Fig.~8.}
\label{m92}
\end{center}
\end{figure}

%---Fig.10
\begin{figure}[!p]
\begin{center}
\includegraphics[scale=0.7]{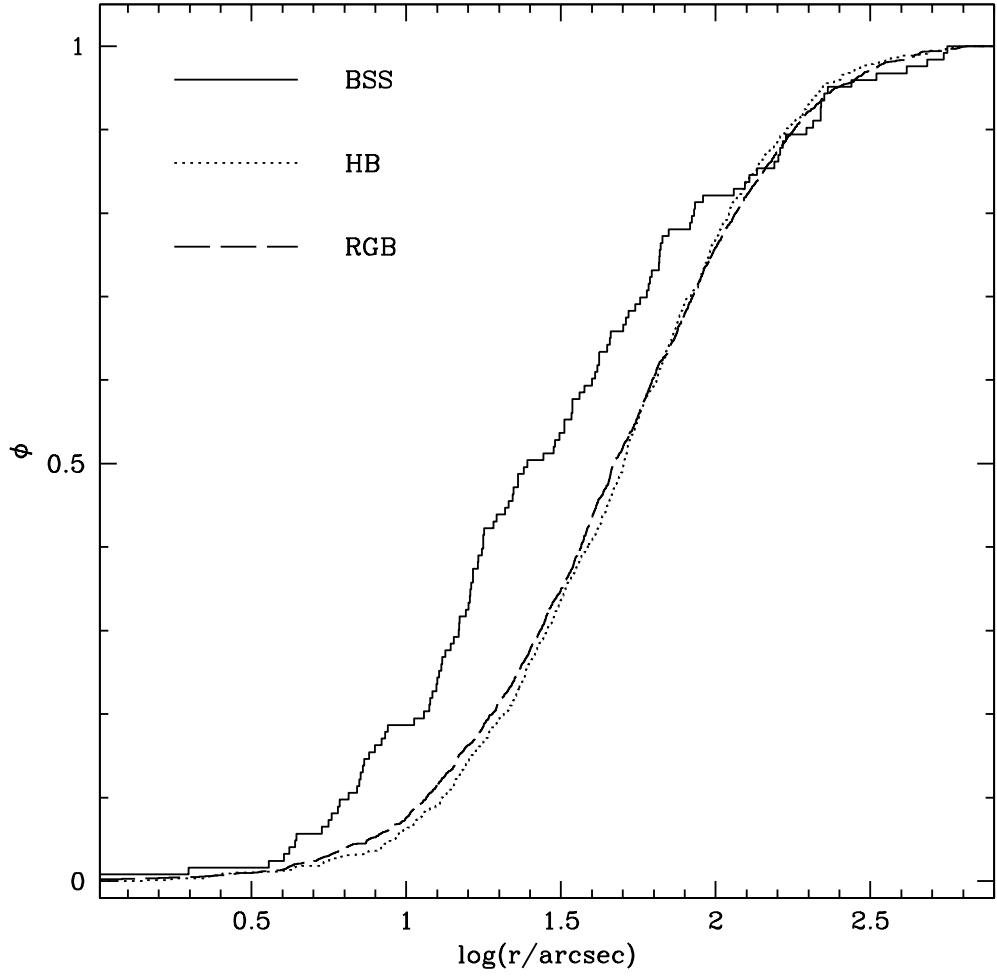}
\caption{Cumulative radial distribution of BSS (solid line), HB (dotted line) and RGB (dashed line) stars as a function
of the projected distance from $C_{\rm grav}$.}
\label{cumrad}
\end{center}
\end{figure}

%---Fig.11
\begin{figure}[!p]
\begin{center}
\includegraphics[scale=0.7]{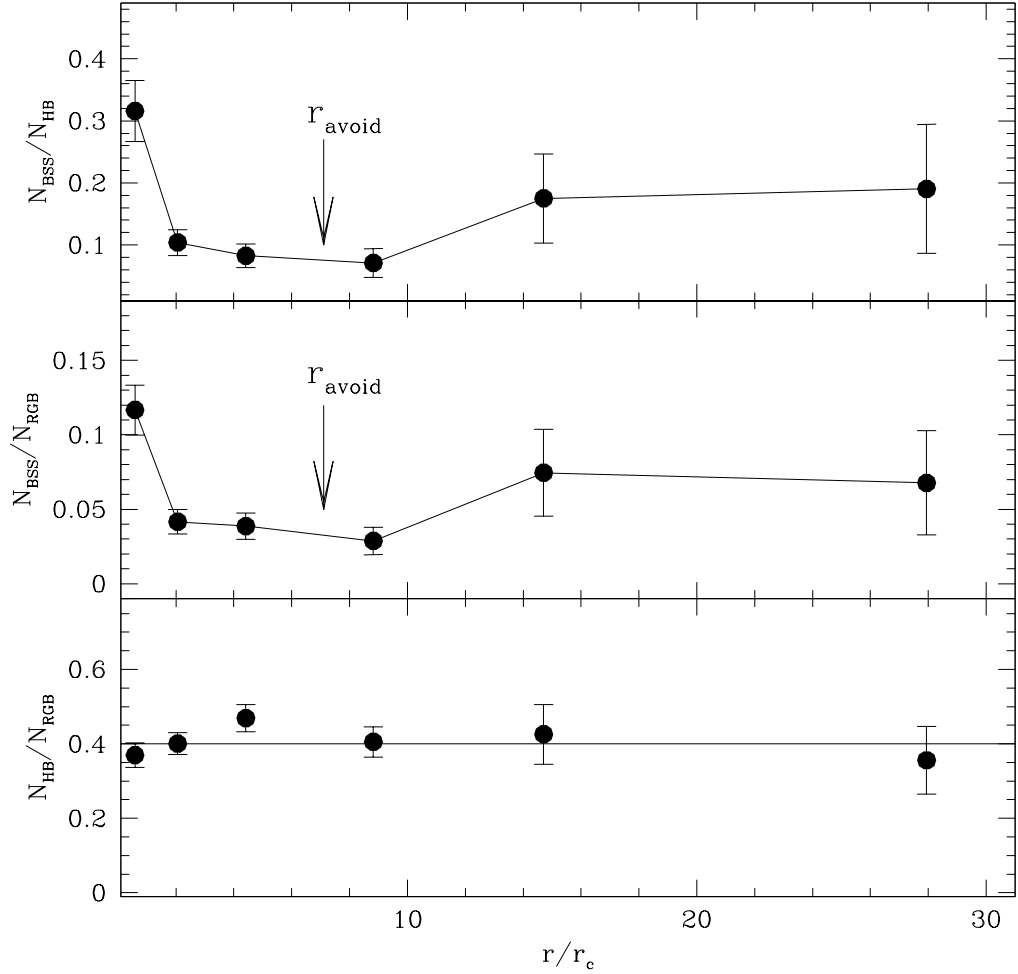}
\caption{Radial distribution of the population ratios $N_{\rm HB}$/$N_{\rm RGB}$, $N_{\rm BSS}$/$N_{\rm HB}$ and
$N_{\rm BSS}$/$N_{\rm RGB}$  as a function of the radial 
distance from the cluster center, expressed in units of the core radius. The arrows mark the position of the radius of
avoidance (see Sect.~5).}
\label{KS}
\end{center}
\end{figure}

%---Fig.12
\begin{figure}[!p]
\begin{center}
\includegraphics[scale=0.7]{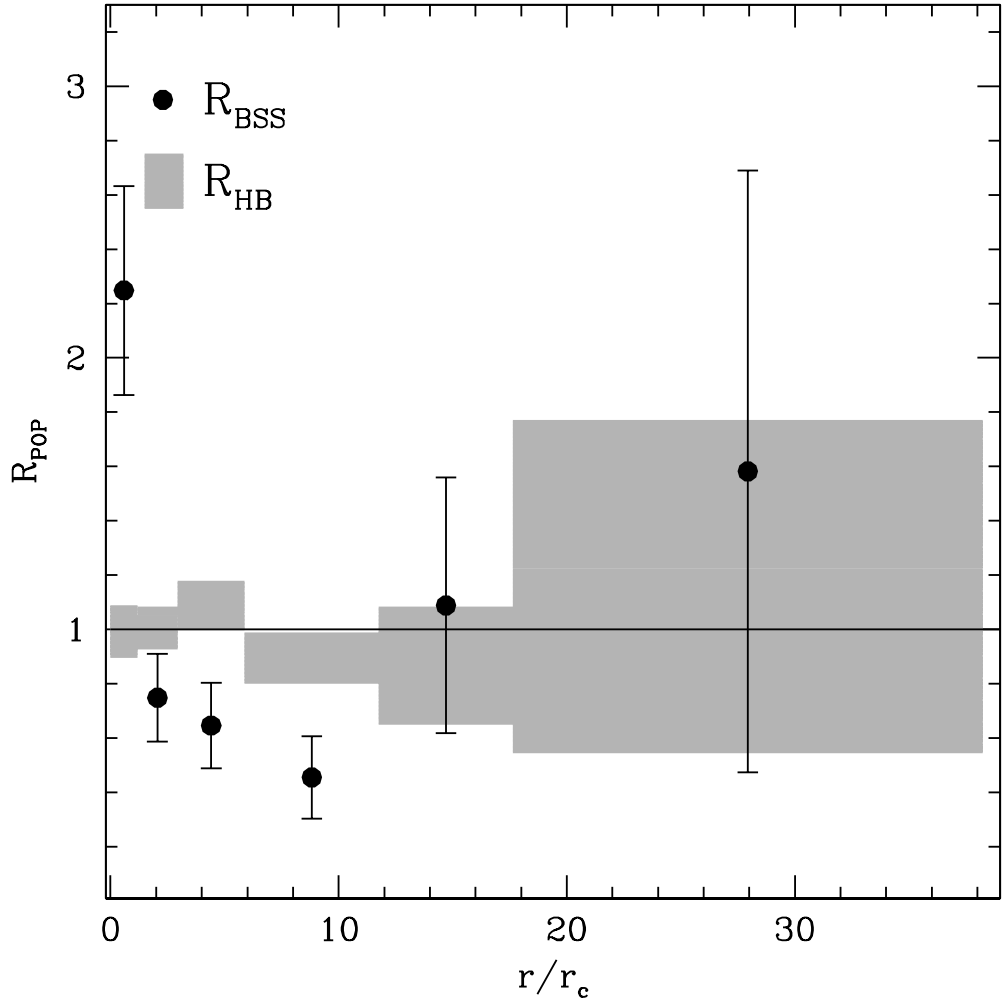}
\caption{Radial distribution of the doubled normalized ratio of BSSs (large dots) and HB 
stars (grey rectangular regions). 
The vertical size
of the grey rectangles correspond to the error bars.}
\label{Rpop}
\end{center}
\end{figure}

%---Fig.9
\begin{figure}[!p]
\begin{center}
\includegraphics[scale=0.7]{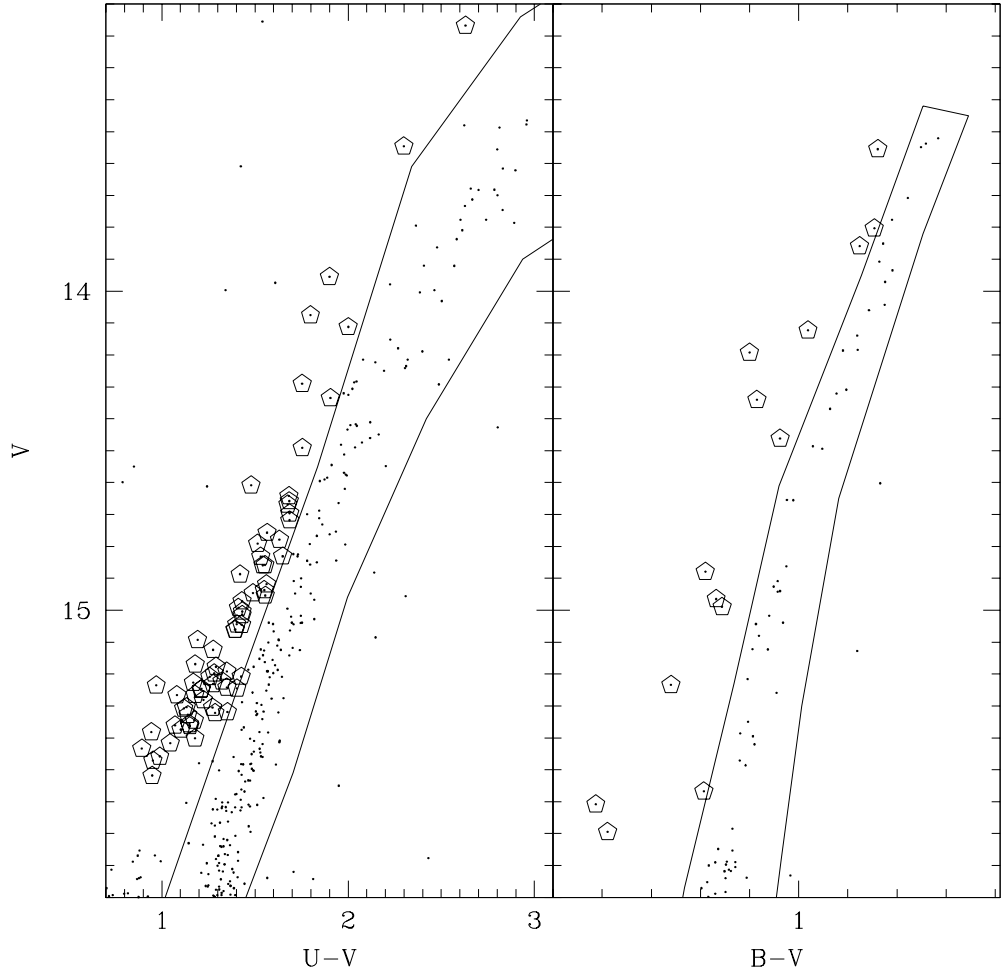}
\caption{Brightest portion of the ($V$, $U-V$) CMD for the WFPC2 sample (left panel) and of
($V$, $B-V$) CMD for the EMMI sample (right panel). The selected AGB stars are marked as pentagons.}
\label{bbisel}
\end{center}
\end{figure}

%---Fig.10
\begin{figure}[!p]
\begin{center}
\includegraphics[scale=0.7]{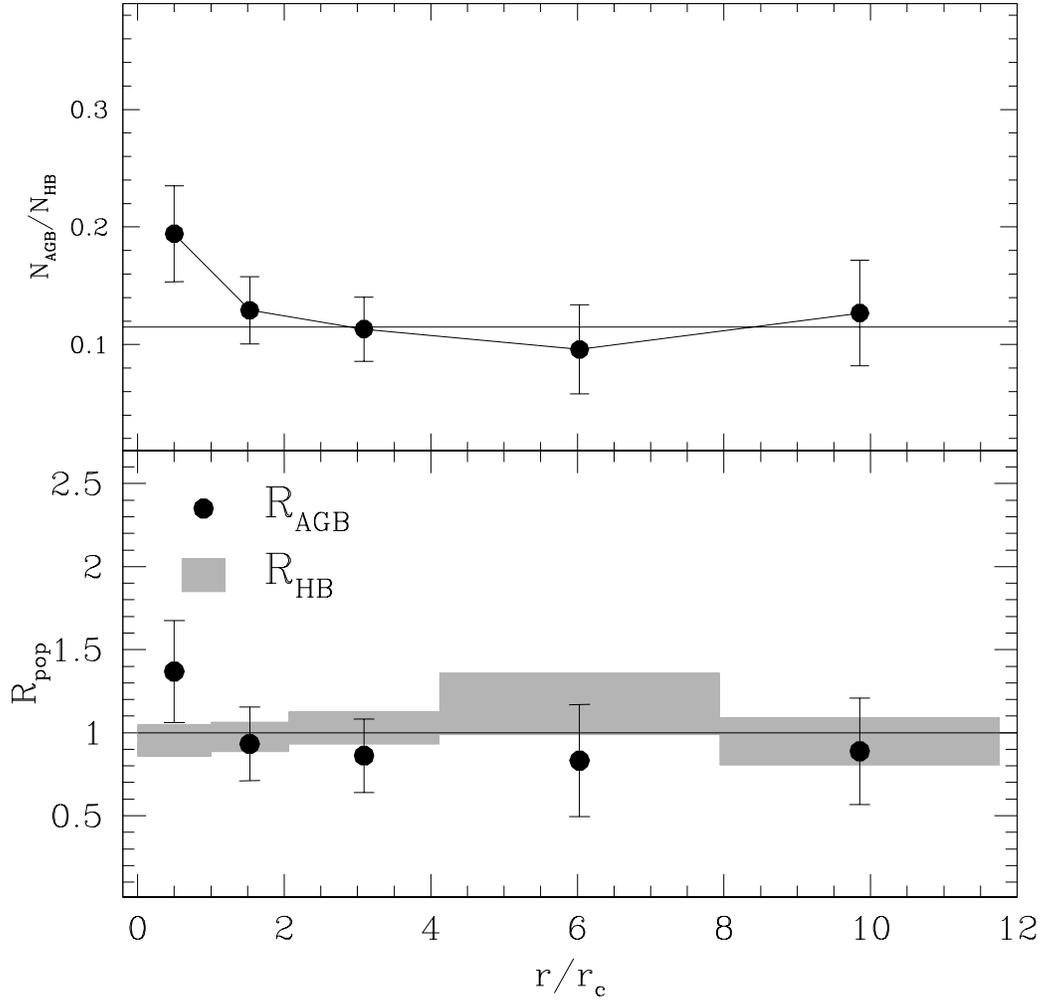}
\caption{Radial distribution of the population ratios $N_{\rm AGB}$/$N_{\rm HB}$  (upper panel) and double
normalized ratio (bottom panel) for AGB (dots) and HB (grey rectangles) as a function of the distance from $C_{\rm grav}$
in units of the core radius. The vertical size
of the grey rectangles corresponds to the error bars.}
\end{center}
\end{figure}

\begin{figure}[!p]
\begin{center}
\includegraphics[scale=0.7]{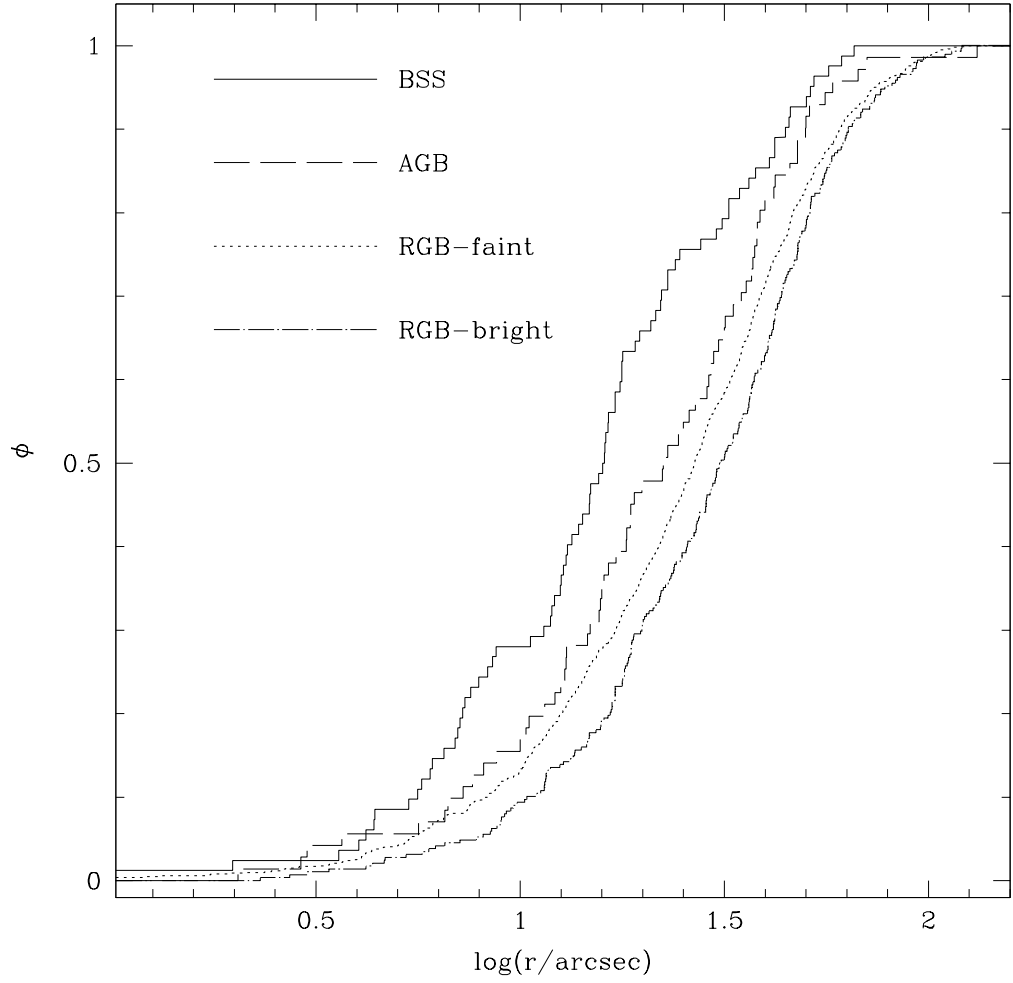}
\caption{Cumulative radial distribution of BSS, AGB, bright-RGB and faint-RGB as selected in the WFPC2
sample.
}
\label{agb-dist}
\end{center}
\end{figure}

\begin{figure}[!p]
\begin{center}
\includegraphics[scale=0.7]{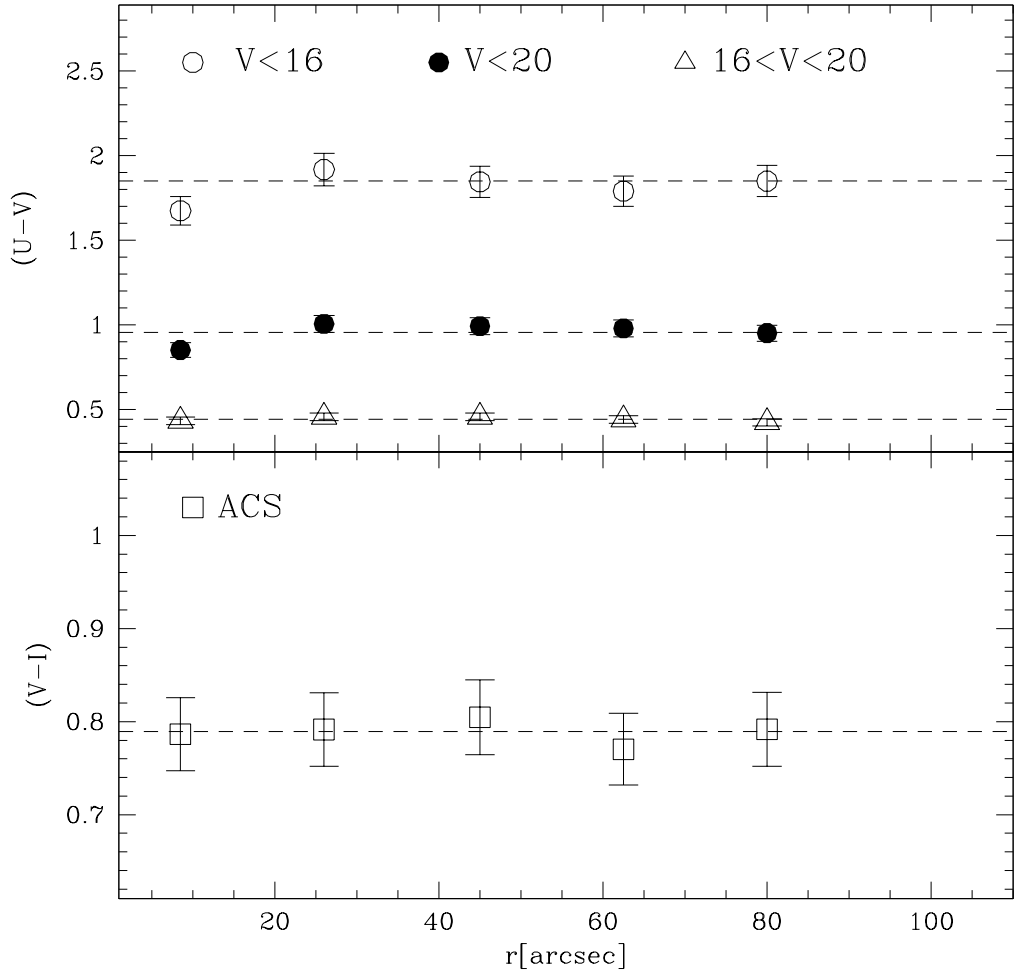}
\caption{Top panel: radial distribution of the ($U-V$) color computed from the WFPC2 resolved stars, for three
different magnitude cuts (see labels). The dashed lines mark the average color computed from the four most
external points. Lower panel: same for the ($V-I$) color computed from the ACS sample.}
\label{col}
\end{center}
\end{figure}

\begin{figure}[!p]
\begin{center}
\includegraphics[scale=0.7]{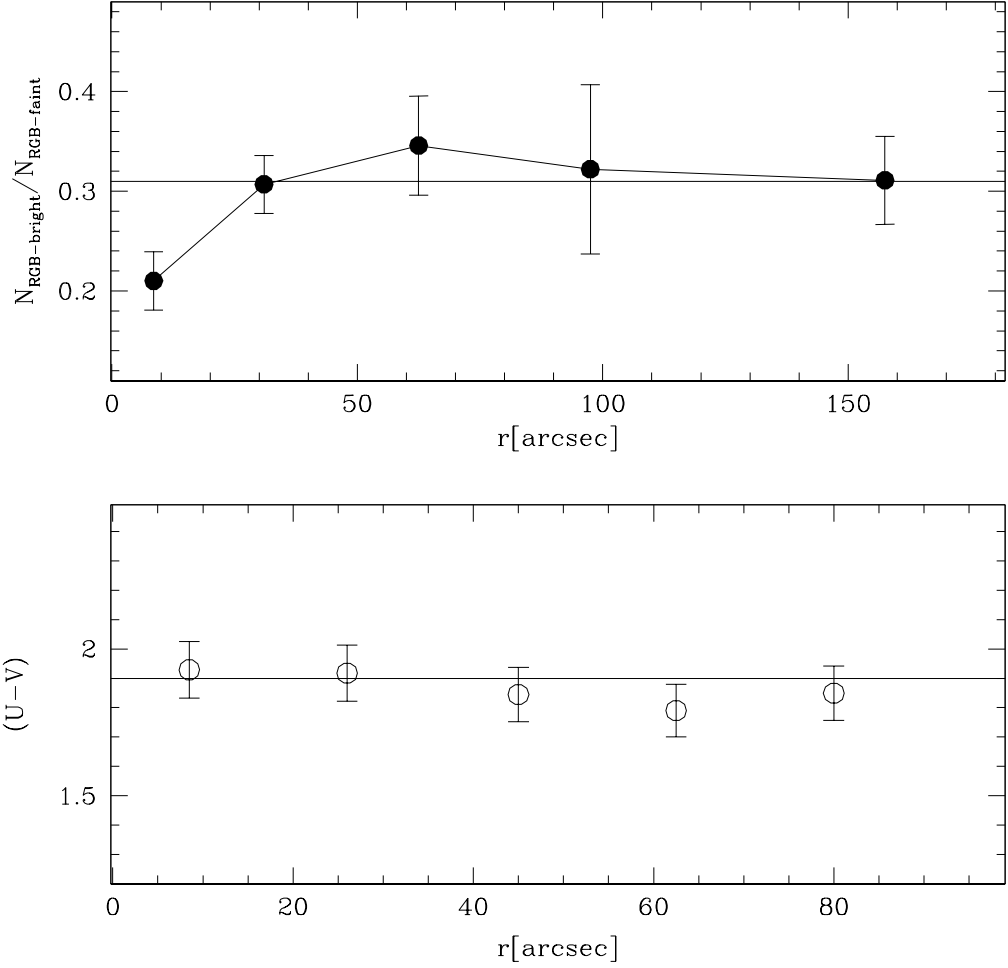}
\caption{Upper panel: radial distribution of the ratio between the number of bright ($V<16$) 
and faint ($V\geq16$) RGB stars in the WFPC2 and EMMI samples. In order to increase the central value of this ratio
to the average one, $\sim25$ stars should be added to the bright RGB population within the cluster core. 
Lower panel: the ($U-V$) color obtained by adopting such an increase is shown panel.}
\label{rgb}
\end{center}
\end{figure}

\end{document}